\documentclass[12pt]{article}
\pdfoutput=1

\usepackage{graphics, color}
\usepackage{graphicx}
\usepackage{amssymb}

\def\gsim{\, \rlap{$>$}{\lower 1.1ex\hbox{$\sim$}}\,}
\def\lsim{\, \rlap{$<$}{\lower 1.1ex\hbox{$\sim$}}\,}

\newcommand{\be}{\begin{equation}}
\newcommand{\ee}{\end{equation}}
\newcommand{\bea}{\begin{eqnarray}}
\newcommand{\eea}{\end{eqnarray}}

\begin{document}


\begin{titlepage}

\setcounter{page}{1} \baselineskip=15.5pt \thispagestyle{empty}

\vbox{\baselineskip14pt
}
{~~~~~~~~~~~~~~~~~~~~~~~~~~~~~~~~~~~~
~~~~~~~~~~~~~~~~~~~~~~~~~~~~~~~~~~
\date{}
}

\bigskip\

\vspace{.5cm}
\begin{center}
{\fontsize{19}{36}\selectfont  \sc
 Inflation in string theory confronts data \\ 
 \vspace{3mm}
  Les mod\`eles d'inflation en th\'eorie des cordes face aux observations
\vspace{2mm}
}
\end{center}


\vspace{0.6cm}

\begin{center}
{\fontsize{13}{30}\selectfont  Eva Silverstein}
\end{center}


\begin{center}
\vskip 8pt
\textsl{
Stanford Institute for Theoretical Physics, Stanford University, Stanford, CA 94306, USA}

\vskip 7pt
\textsl{ SLAC National Accelerator Laboratory, 2575 Sand Hill, Menlo Park, CA 94025}



\end{center}

\vspace{0.2cm}
\hrule \vspace{0.1cm}
{ \noindent \textbf{Abstract} \\[0.2cm]
\noindent 

{Following the 2015 {\it Planck} release, we briefly comment on the status and some ongoing opportunities in the interface between inflationary cosmology, string theory, and CMB data.  The constraints in the $r$-$n_s$ plane introduce a new parameter into inflationary cosmology relative to the simplest quadratic inflation model, in a direction which fits well with couplings to heavy fields as occurs in string theory.  The precision of the data permits further searches for and constraints on additional model-dependent features, such as oscillatory $N$-spectra, a program requiring specific theoretically motivated shapes.  Since the perturbations can easily be affected by additional sectors and couplings, null results can usefully bound such contributions.  We also review the broader lessons string theory has contributed to our understanding of primordial inflation, and close with some approaches to a more complete framework.  Published in a special volume of Comptes Rendus on {\it Inflation:  Theoretical and Observational Status}.       
}
}

 \vspace{0.3cm}
 \hrule

\vspace{0.6cm}
\end{titlepage}

\tableofcontents

\newpage

\baselineskip = 16pt

\section{Introduction}

 With the {\it Planck} 2015 release, the highly nontrivial agreement between $\Lambda$CDM and the CMB remains robust \cite{PlanckParameters}\cite{PlanckNG}\cite{BKP}\cite{PlanckInflation},  reproduced via an independent set of tests in polarization.  Inflationary cosmology fits this data extremely well, predicting that the power spectrum (which is {\it a priori} a general function) boils down to a few numbers to a good approximation.  This is a spectacular development, giving an account of the origin of structure which follows from the basics of quantum mechanics combined with accelerated expansion of the early universe \cite{classicperts}.  At the same time, the subject is rich with interesting unanswered questions, both phenomenological and beyond -- including the physics of earlier phases of the universe and the search for  a complete framework for describing formal `observables' in cosmology.  

Moreover,  although inflation provides a paradigm which can be described using familiar quantum field theory and general relativity, quantum gravity effects play a role:  any low-energy model of inflation makes assumptions about Planck-suppressed operators.  This offers remarkable contact between string theory, low energy effective theory, and observables.  More generally, the observable perturbations can easily be affected by additional sectors of fields (or higher dimensional defects), either heavy or light compared to Hubble during inflation, as well as being sensitive to nonlinear interactions in the inflation sector.  As a result, either discoveries or constraints can be very informative for high energy physics.  

The goal of understanding the underlying principles and mathematical structures in physics suffices to justify substantial effort in formal theory \cite{Hardy}, and many important questions will always remain out of empirical reach.  However, the interface with cosmology provides an overlap between thought experiments and real ones which is well worth exploiting to the full extent possible.  From the point of view of string theory research, the formal questions in cosmology are certainly as interesting as in other areas, with the added benefit of potential connections with data given the relatively high energy scales and field strengths that may be involved in inflation.    

Current observations provide powerful constraints and discovery potential along several axes of theoretical and phenomenological interest, getting closer to important thresholds for tensor modes and various non-Gaussianity parameters.  The data provides opportunities for additional searches for structures in the power spectrum or higher correlators.   In this note,\footnote{For more comprehensive introductions and reviews see e.g. \cite{DanielLiamBook}\cite{LesHouches}\cite{CliffTalk}}  I will provide some updated comments on this interface and propose some next steps following the 2015 Planck release along with the progress in B mode detection by a number of experiments as well as new approaches in large scale structure.     

\section{$r$ and $n_s$:  Heavy field(s), deformations of $m^2\phi^2$, and new cosmological parameter(s)}

Although $\Lambda$CDM fits the data beautifully, there is already a sense in which we require an additional parameter given the latest CMB results. 

%

Including the analysis \cite{BKP}, the paper \cite{PlanckInflation}\ reports a reduction by 29 percent of the viable region of the $r-n_s$ plane, enough to further disfavor $m^2\phi^2$ inflation \cite{Andreichaotic}\ (as well as Natural Inflation \cite{Natural}).  Let us proceed under the assumption that these scenarios are in fact ruled out, although this may be premature. The $m^2\phi^2$ model incorporates both the inflationary phase with the resulting density perturbations and the exit from inflation that is required to match to later-time cosmology,  with just two nonzero parameters: $m$ and the initial position of the field.   Since inflation must exit, and assuming analyticity in field space, there must be one or more additional parameters in the inflaton action as compared to this model.  Although the 6 parameters of $\Lambda$CDM suffice to characterize CMB data, the theory now requires an additional parameter in order to capture that combined with the empirical fact that inflation must end.   There are many ways such a new parameter could enter.  

The direction it goes fits very well with basic expectations from the point of view of UV completing gravity \cite{flattening}:  adjustments of heavy fields involved in the UV completion tend to flatten the potential since that is energetically favorable.    
It is worth emphasizing that this might be expected in any UV completion of gravity, since new degrees of freedom  would be expected to take over at or below the Planck scale where the gravitational coupling becomes strong.   It was articulated originally in the context of string theory, which is arguably
the leading candidate\footnote{There are interesting proposals for alternative frameworks for quantum gravity including inflation, such as \cite{BanksFischler}.} for a theory of quantum gravity which has passed numerous thought-experimental `null tests'.\footnote{These include calculations of black hole entropy, duality relations which make sense of strong-coupling limits (including the AdS/CFT correspondence and other formulations of quantum field theories), a plethora of evidently metastable local minima of the scalar potential, permitting an interpretation of the cosmological constant as a selection effect, as well as the present topic of inflationary mechanisms which account for quantum gravity effects and fit well with current data.}

String theory contains various scenarios for inflation.  One basic mechanism, monodromy inflation \cite{MonodromyI, MSW, FMPWX, ignoble, unwinding, recentmonodromy, powers}\ , realizes large-field (`chaotic') inflation generalizing the $m^2\phi^2$ model \cite{Andreichaotic} (with elements of natural inflation \cite{Natural}).  This mechanism descends in a robust way from higher dimensional gauge potentials and their couplings to magnetic fluxes, as well as various analogues related by string duality symmetries.  Examples of this mechanism predict potentials flattened \cite{flattening}\cite{powers}\cite{recentmonodromy}\ compared to fiducial monomial forms $\phi^n$  (and with small oscillatory features derived from the underlying periodicity of the theory in axionic directions).  For $n=2$, the adjustments of heavy fields typically produce a potential of the form $\phi^{p<2}$ (for the single field version of the mechanism).  This fits within the currently allowed region in the $r-n_s$ plane, as a direct result of the flattening effect. 

This effect is just simple classical mechanics at the level of the potential $V(\{\phi_{H}\}, \phi)$, incorporating its  dependence on both the candidate inflaton $\phi$ and heavy fields $\{\phi_H\}$.  Note that another energetically favorable possibility is simply that the inflationary energy destabilize the potential in some direction in field space.  To assess this it is useful to bring in some basic elements of string theory compactifications.  The simplest structures entering in moduli stabilization are three-term or two-term stabilization mechanisms.  The latter come from a balance of fluxes (magnetic fields) threading through different directions in the extra dimensions.  The flux prevents either direction from shrinking too much at fixed overall volume, since the energy density in the magnetic field would blow up.  This  leads to a potential of the schematic form 
\begin{equation}\label{schematicf}
f_1(\phi)x^{\gamma_1}+f_2(\phi)x^{-\gamma_2}
\end{equation}
(with $f_1, f_2, \gamma_1,\gamma_2>0$), where $x$ is a ratio of length scales in the two different directions.  At fixed $y$ these two terms stabilize $x$.  Because of the Stuckelberg structure of the gauge field kinetic terms, one obtains a potential depending on the flux quanta $Q$ and axions $\phi$ for which the terms in (\ref{schematicf}) are schematically of the form 
\begin{equation}\label{fluxax}
(Q_1+ \frac{\phi}{f} Q_2)^2 x^\gamma
\end{equation}
As a result, the potential fiducially depends on the inflaton $\phi$ as $\sim \phi^2$, and the adjustment of $x$ flattens the potential relative to this.\footnote{Higher fiducial powers are also possible \cite{recentmonodromy}\cite{powers}.}  In contrast, 
three-term stabilization mechanisms, with coefficients depending similarly on $\phi$, do not tolerate large field ranges before getting destabilized at least for modest values of discrete parameters.  It is therefore the former mechanism which operates in concrete models proposed in the literature.  Couplings in the kinetic term similarly affect the potential for the canonically normalized field.  These can go in either direction (flattening or steepening), something that would be interesting to characterize more systematically.  
        
The predictions for the tensor to scalar ratio for the earliest examples of this mechanism are very close to the current peak value of the posterior on $r$ in the BICEP-Keck-Planck joint analysis \cite{BKP}, but the latter is also  consistent with $r=0$, as is all current data at the present writing.       
The constraints on $n_s$ are also interesting, although significantly more fungible in these models including the possibility of multiple fields, which reddens the tilt \cite{Danjie}.  I will discuss oscillatory features, non-Gaussianities, and related effects separately below.   The observational progress on the tensor to scalar ratio $r$ is extremely interesting in its connection to quantum gravity, enabling a measurement or constraint on quantum-gravitational waves and on the field range relative to the Planck scale during inflation \cite{Lyth}.  

It will be interesting to improve the theoretical constraints on model parameters as the observational constraints continue to tighten.  One basic question is whether it has simply been a coincidence that the predictions originally landed within the viable region of $r$ and  $n_s$ and remain so at the present writing.  Starting from the quadratic model, we had explained this improvement of fit in a simple way above via the energetically favorable flattening effect \cite{flattening}.  But as described in \cite{powers}\ and \cite{recentmonodromy}, there exist couplings in the theory which admit a starting (un-backreacted) power $V\propto \phi^n$ with higher $n$.  In fact,  $n$ is in principle unbounded, taking into account that perturbative string theory can be formulated in any total dimensionality $D$, as long as one properly accounts for the tree level potential energy $\propto (D-10)g_{eff}^2$, with $g_{eff}$ the four-dimensional effective string coupling.  These models are (so far) less well studied, and involve additional fields which may have unrealistic consequences, but they raise the question of how large the theoretically-allowed region of $r,n_s$ is in this general mechanism.      

One approach to this is to push the theory toward putatively large values of the fiducial power $p_0=n$, to see whether the kinetic terms and  potential adjustments generally conspire to lower the power to a viable regime.  There is some evidence for this \cite{SCproject}:  for example, considering twisted tori in any dimension, generalizing 
\cite{twistedtori}\cite{MonodromyI}\ yields at most a finely spaced `discretuum' of models  with $p<2$.

Another direction of current interest is the relation of multi-field Natural Inflation \cite{multiax}\ (where the monodromy effect is turned off, but multiple axion fields are included) to the `weak gravity conjecture' \cite{WGC} \cite{LiamMulti} \cite{otherWGC}.  Theoretical constraints on the charge to mass ratio of black holes are related to certain instanton effects which can reduce the period and hence field range {\it if} their amplitude is sufficiently large.  However, at least at the present writing there is no evidence for a universal constraint on this mechanism \cite{LiamMulti}.  In monodromy inflation, the value of the underlying axion period $f$ is not relevant for the field range; a classical potential lifts the axion, and residual periodic effects are subdominant, leading to small oscillations in the potential and periodic particle or string production effects as we will discuss further below.  At large radius and weak coupling, perturbation theory is a good approximation and nonperturbative effects such as instantons are exponentially suppressed.  Much stronger conditions have been considered, for example in \cite{ReeceEFTcondition}, which noted that in axion monodromy one does not have the same light spectrum all along the field range.  This is well known, already true in simpler, less realistic systems with large field ranges in string theory, such as supersymmetric gauge theory moduli spaces.  In the inflationary context, this arises in two ways: (i) from the monodromy of wrapped branes of all dimensions in the theory, leading to a new set of lightest degrees of freedom each time around the underlying period, and (ii) an increasing number of light modes carried by the monodromy-inducing brane/flux sector.              
Concrete models address these effects, and no physical principle forbids them; they do not generally present complications at the Planck field range (or well beyond).    

There are other string-theoretic mechanisms for slow roll inflation which predict small $r$ and viable $n_s$ (depending on details), such as the pioneering work \cite{KKLMMT}\ (with undetectably small $r$) and proposals to realize, among other things,  Starobinsky-like inflation with exponential terms in the potential \cite{stringStarobinsky}\cite{Roulette}\ (with potentially detectable $r\sim 10^{-3}$).  This can also arise naturally from extra-dimensional geometry in some regimes (see the talk by C. Burgess \cite{CliffTalk}\ for a recent review).  It would be interesting to understand more systematically the corrections in Starobinsky inflation \cite{DanielLiamBook}.  Other interesting deformations studied recently in a supergravity framework include parameters $\alpha$ in the kinetic terms, related to the geometry of scalar field space \cite{KalloshLinde}\ (covered in another chapter in this volume by A. Linde).    
It would be interesting to understand how these might be connected to UV physics.

\section{Comments on the nature of observational predictions in string theory}

It is notoriously difficult, if not impossible, to produce useful global predictions from string theory.  On the one hand, the theory is {\it highly} constrained -- the absence of a hard cosmological constant in any dimension, and the sub-Planckian range of single axion fields (without incorporating monodromy) being two basic examples relevant to cosmology. 
The theory codifies basic principles such as unitarity, ultraviolet compoleteness, black hole statistical mechanics and holography. On the other hand, there are many solutions of string theory.\footnote{This latter property can in itself be considered a feature rather than a bug, as it lines up with a viable interpretation of the dark energy as a selection effect.}

This is quite analogous to quantum mechanics or quantum field theory models of physical phenomena.   General principles such as unitary evolution are very important and powerful, but do not in themselves provide specific predictions. Such predictions only arise after specifying the degrees of freedom, Hamiltonian, and wavefunction of the particular model or class of models. The difference in inflationary cosmology is that understanding the process requires control of quantum gravity effects.  So one models them ideally in a quantum gravity theory, in particular string theory. 

In that context, similarly to quantum mechanics or QFT, specific predictions arise only after further specifications.  Mechanisms and models have been identified in this way, which take into account structures and constraints in string theory.  Those which produce detectable $r$ (or don't) and those which predict interesting shapes of non-Gaussianity (or don't), are very different microscopically.  Being able to cleanly distinguish different models in quantum gravity, in a field which requires control of such effects, is groundbreaking (thanks to the monumental observational progress in the field). This is true  despite the absence (at least at the present writing) of a universal prediction.  

Independently of the observational constraints, another application of string-theoretic inflation has been to illustrate very concretely the full range of inflationary dynamics, feeding into more systematic effective field theory (EFT) treatments. 
Even if in the end the date hones in on single-field slow roll inflation with undetectably small $r$, this more accurate understanding of the dynamic range of inflation is important.       

That said, there is room for progress in the direction of more powerful no go theorems, or more general, less model-dependent string theoretic effects (for example exploiting the intrinsic non-localities arising from its extended objects \cite{spreading}, or the pattern of massive states arising in string theory \cite{juannima}).  Regardless, it is very important not to generalize prematurely from a small set of examples or conjectures.  The ultimate lessons will reveal themselves on an appropriate timescale, and patience is warranted both theoretically and experimentally.

\section{Structures in the power spectrum and non-Gaussianity}

Although the the data favors $\Lambda$CDM over all alternatives so far analyzed -- and does so in a highly nontrivial, overconstrained way -- it still leaves room in principle for additional structures.  Once again for both the power spectrum and the higher point correlators the {\it Planck} 2015 results impose important constraints \cite{PlanckParameters}\cite{PlanckInflation}\cite{PlanckNG}, finding
no statistically significant evidence for such additional structures\footnote{although Planck reports a set of  $~3 \sigma$ anomalies in multifrequency oscillatory non-Gaussianity and at low $\ell$ in the power spectrum}.  But it remains an open possibility worth exploring in the spirit of making the most of the data, particularly since various mechanisms and symmetry structures lead to ancillary predictions along these lines, analyzed in interesting work such as \cite{oscillationsdata}\cite{Munchmeyer}\cite{EFToscillations}.   Turning this around, tightening constraints can put very interesting limits on inflaton couplings and additional sectors \cite{PlanckSuppressed}.   

This is a place where the `UV completion' can play a special role.  Although from a low energy effective field theory point of view \cite{PofX}\cite{EFTetc}, one can beautifully organize the observables \cite{classicperts}, there are arbitrary functions of time $t$ in the resulting Lagrangian (without additional symmetry assumptions beyond what is required for inflation \cite{EFToscillations}).  Even if some additional structure were present, searching for such arbitrary functions (with many parameters) could never provide a significant detection.  To make progress in constraining such possibilities therefore requires theoretically motivated structures with a limited number of parameters. 

Broadly speaking, mechanisms that produce inflation via nontrivial interactions which slow the inflaton tend to produce strong non-Gaussianity at the single-clock level, leading to overlap with the equilateral and flattened shapes for the bispectrum.  An early example of this was DBI inflation \cite{DBI} and its cousin trapped inflation \cite{trapped} (which is related to both DBI inflation and monodromy inflation by varying parameters, with some interesting intermediate regimes where all may play a role as in the Unwinding Inflation scenario \cite{unwinding}).   From the bottom up, at the single-field level these obey a model-independent relation to the sound speed of perturbations \cite{EFTetc}.    
Some mechanisms contain an underlying periodic structure, leading to oscillatory features; in general features can arise in a variety of ways via interactions of the inflaton with additional degrees of freedom.  With multiple fields, the possibilities proliferate.    

The {\it Planck} team analyzed a plethora of non- Gaussian shapes, including templates combining equilateral and flattened shapes with features or oscillations.  When allowing for multiple frequencies, some such combinations grew in significance with the inclusion of polarization, some exceeding $3\sigma$ after look elsewhere effects were taken into account according to the analysis of \cite{PlanckNG}.  
The $\gtrsim 3\sigma$ improvements could well be statistical fluctuations, or as yet unresolved systematics, consistent with the underlying model being simple slow roll inflation.  Nonetheless, it is interesting to take a phenomenological approach to the problem and ask whether an appropriate theoretical structure could underlie the templates which yielded the greatest significance in the data -- and if so whether the structure might tie together the parameters in a way that substantially improves the significance and/or makes further testable predictions.  Let me make a few comments along these lines, based on preliminary discussions \cite{partprodfeatures}; see also \cite{DanGreen}\ for another approach.

From the string-theoretic point of view, logarithmic-spaced oscillations arise automatically from the structure underlying axion monodromy, with model dependent amplitude.    These periodic effects in general include both an instanton-generated sinusoidal term in the potential and also particle or string production events of various types.  The former have been analyzed extensively, but the latter are also as intrinsic to the mechanism.  They have been studied in more extreme regimes of parameters \cite{trapped}, where they provide a novel mechanism for inflation with strongly non-Gaussian perturbations.  In that regime, the perturbations were studied in a continuum approximation which washed out the oscillations.  
It will be interesting to instead work in an essentially slow roll inflation regime and relax the continuum approximation, incorporating the production-induced oscillations and deriving the resulting amplitudes and shapes \cite{partprodfeatures}\ of non-Gaussianity and its relation to additional features in the power spectrum.   

Finally, it is worth noting that although the CMB data has provided useful constraints, it cannot reach some important thresholds -- in particular the thresholds separating slow roll inflation from alternative mechanisms \cite{EFTetc}.  These will require additional data such as large-scale structure \cite{SphereX}, which may ultimately provide additional constraints on new physics.



\subsection{Multifield dynamics and statistics}

There has been a lot of interesting recent work on multifield effects, and also on statistics of multiple vacua in certain classes of effective actions.  The presence of multiple light (or temporarily light) fields brings in a variety of additional phenomena.  In addition to enriching the basic geometry of the field trajectories, non-adiabatic and renormalization effects arise with back reaction on the inflaton motion and perturbations.  This is a complicated problem in general, but some very interesting results have been obtain in particular circumstances.  Some recent examples include \cite{multifield}\cite{DanGreen}.

\section{Signatures of defects (bubbles, strings, etc.)}

In some sense, the most striking potential observables related to string theory might arise from defects such as bubble walls (and/or the negative curvature inside a bubble), or cosmic strings, or even inflation at such a high scale compared to the string tension that string-theoretic effects become directly detectable \cite{juannima}.  

The observation of bubble nucleation or collision requires special conditions -- a small enough number $N_e\approx 50$ of e-foldings of inflation to preserve evidence of the onset of inflation, i.e. just enough to be phenomenologically viable.   In an interesting class of examples based on the KKLMMT mechanism \cite{KKLMMT},  with an accidental cancellation among several hundred operator contributions to inflation, the distribution of e-foldings was shown to decrease at large $N_e$ \cite{LiamNe}.  This is related to the percent level tuning required in this type of mechanism for small-field inflation.  

However, it is very important to note that inflation need not be fine-tuned from the point of view of the standard, Wilson-'tHooft naturalness criteria.  With a shift symmetry, corrections do not ruin slow roll, and the hierarchy between the scale of the potential and the Planck scale can easily be obtained via dimensional transmutation (as well as its various duals).     These criteria embed consistently in string theory given the discrete shift symmetries along axion directions, given proper adjustments of fields which couple to them such as the `moduli' scalars.  (There may be other Wilsonian-natural regimes, but this is enough to suggest a very large class of examples with no reason for a minimal $N_e$.)  For this reason, I know of no non-circular arguments for a minimal number of e-folds of inflation in the string landscape.  
Regardless, this is a very interesting class of observables to pursue \cite{bubbles}.  
\footnote{The low power at low $\ell$, another non-statistically significant potential hint in the data can be interpreted in terms of this hypothesis.  Other approaches to that include a wide initial oscillation, perhaps correlated with higher-$\ell$ physics  (see some of the references in \cite{oscillationsdata}).  However, given that the dust contribution to \cite{BKP}\ is at least half the signal, there is less significant evidence favoring such modifications than would have been the case otherwise.}  

Cosmic strings may be produced during inflation, influencing the inflaton evolution and perturbations \cite{exoticsources}, or during the exit \cite{joereview}, remaining potentially visible in the late universe.  
There is no evidence for such sources currently, but again further searches will be possible with additional data.  
Reheating is generally an interesting phase that is not completely mapped out, with some interesting connections to string inflation as well (see e.g. \cite{reheatingegs}).  

\section{Null results and {\it Planck}-suppressed operators}

So far in this note we focused on the rapidly developing observations and related opportunities for connections to string theory.  Let us finally discuss the possibility that the data will reveal no new parameters (aside from the one argued for above).  This may or may not be the case, but it is  worth asking the question:  if tensor modes above $r\sim .01-.001$, non-Gaussianity above the slow-roll threshold, and additional structures are all eventually excluded to the extent possible in the data, what would we have learned from this approach to early universe cosmology?  

$\bullet$  Since the perturbations can so easily be infected by additional sectors and couplings to give effects beyond $\Lambda CDM$, one can turn it around and place interesting bounds on additional sectors and couplings (see e.g. \cite{PlanckSuppressed}\ for some examples).  

$\bullet$  The sensivitity to Planck suppressed operators strictly speaking requires control over quantum gravity, so the developments showing how inflation very plausibly fits into string theory naturally in a variety of ways has been important \cite{DanielLiamBook}.  The string theoretic mechanisms studied thus far by and large remain viable, and some would remain so if all results are null.    

$\bullet$  String theory taught us interesting lessons about how broadly inflation works.
A null result is basically as interesting as what it excludes, so learning via concrete mechanisms that inflation could be large-field or small, that it could essentially Gaussian or not, is worthwhile.    Then deciding among these classes in our observed universe is useful, especially since they are so different microscopically.

$\bullet$  Relatedly, the `top down' approach feeds directly into the effective field theory analysis and some data searches.  It has greatly clarified the role of high energy fields and quantum gravity, and helped expose some of the limitations of folk theorems in the subject. 

$\bullet$  Finally, at a purely theoretical level it has taught us some interesting lessons about time-dependent  quantum field theory and string theory.  One example is the speed limit on scalar field evolution in certain strongly coupled large-N gauge theories \cite{DBI}, and the interpolation between this and weak-coupling particle production effects \cite{trapped}.    From the point of view of of formal theory, the subject is as interesting as any other set of topics in string theory, but with the added chance for further connection to observational data.

\section{Toward a framework}

Athough the phenomenology of inflation works very well, there is not a complete theoretical framework for cosmology, something that is challenging for a variety of reasons.  String theory is massively dominated by cosmological solutions -- almost every choice of dimensionality and topology of the extra dimensions leads to a leading positive term in the scalar potential energy.\footnote{This may not be familiar, as much early work in string theory involved setting this to zero by hand, working in the critical dimension with Ricci-flat extra dimensions.}  However, its flat and {\it anti-}de Sitter solutions, while unrepresentative, are far simpler to analyze systematically.  This
has led to a complete framework for quantum gravity in anti-de Sitter spacetime and its relevant deformations \cite{AdSCFT}, with the observables defined as those of a quantum field theory (QFT), with the warping (gravitational redshift) in the gravity solution corresponding to the energy scale in the QFT. 

A straightforward method to upgrade this to a partial formulation of de Sitter and other cosmological backgrounds is to start from an $AdS/CFT$ solution and add ingredients which produce a cosmological solution instead, determining their effect on the `brane construction' from which the dual QFT description arises.  The de Sitter observer patch constitutes a warped geometry with {\it two} low-energy (i.e. strongly gravitationally redshifted) regions, but without arbitrarily high energy regions.  This suggests a dual description in terms of two low energy theories cut off at a finite scale in the UV, and coupled to gravity in one lower dimension.  As explained in \cite{micromanaging}\cite{LesHouches}, this is precisely what happens to the brane construction.  Most interestingly, this is a direct result of the structure of runaway directions in moduli space in string theory, which does not admit a hard positive cosmological constant.   The non-perturbative decay of de Sitter spacetime leads to a description ultimately decoupled from gravity. This duality is known as the dS/dS correspondence.  Another approach, known as the dS/CFT correpondence \cite{dSCFT}, has proved useful in characterizing the aspects of observables (including phenomenological ones) that descend from the symmetries of global de Sitter spacetime.  However, it is less clear how to obtain this (at large radius) from an ultraviolet-complete theory, although a very interesting intrinsically small-radius example is given in \cite{higherspin}.  
For another perspective on this problem see \cite{Dioreview}.  

\section{Final remarks}

The coming years promise to be exciting, with a crucial range of $r$ to be covered observationally along with further opportunities for new physics 
searches in CMB and large scale structure data.   It will be interesting to fully flesh out the phenomenology of well-defined inflationary mechanisms and test for their signatures, in the spirit of making the most of this precious resource.  Theoretically we are driven by major puzzles in cosmology and beyond, with hints of new effects that in turn might lead to additional opportunities for contact with observations.  In any case, the characterization of the primordial seeds for structure remains a spectacular development, with the added benefit of a connection to quantum gravity.        

{\bf Acknowledgements} {This note covers only a few aspects of a large field to which many have contributed.  I would like to thank all my collaborators in this subject (along with many others) for sharing their insights and for the pleasure of their camaraderie in the pursuit of these questions.  For the present note I am especially indebted to J. R. Bond, F. Bouchet, S. Gratton, G. Efstathiou, R. Flauger, D. Meerburg, H. Peiris, and B. Wandelt, among others, for discussions of the {\it Planck} 2015 results, and members of the BICEP/Keck, Spider, ACTPol, SPTpol, and Litebird teams for explaining various aspects of the ongoing program of B-mode measurements.  Most importantly, I am grateful to all of those involved for these extraordinary observational developments which so powerfully constrain our understanding of the primordial universe.}

\end{document}